\documentclass{article}
\usepackage{amssymb}
\usepackage{amsmath}
\usepackage{amsfonts}

\newtheorem{theorem}{Theorem}

\newtheorem{proposition}[theorem]{Proposition}

\input{tcilatex}
\begin{document}

\title{Quasi-Periodic Solutions of (3+1) Generalized BKP Equation By Using
Riemann Theta Functions}
\author{ Se\c{c}il Demiray$^{a}$\thanks{{\footnotesize Corresponding Author.
\ Tel.: \ +90 228 214 16 81; \ \ \ \ \ \ \ E-mail address:
secil.demiray@bilecik.edu.tr }} ,$^{b}$Filiz Ta\c{s}can G\"{u}ney \\
$^{a}${\footnotesize \ Bilecik Seyh Edebali University, Bozuyuk Vocational
School,}\\
{\footnotesize \ Bilecik-TURKEY\ \ }\\
{\footnotesize \ \ \ \ \ \ \ \ \ \ \ \ \ \ \ \ \ \ \ \ \ \ \ \ \ \ \ \ \ \ \
\ \ \ \ \ \ \ \ \ \ \ \ \ \ \ \ \ \ \ \ \ \ \ \ \ \ \ \ \ \ \ \ \ \ \ \ \ \
\ }\\
$^{b}${\footnotesize Eski\c{s}ehir Osmangazi University, Art-Science
Faculty, }\\
{\footnotesize Department of Mathematics-Computer, Eski\c{s}ehir-TURKEY\ }\\
{\footnotesize \ \ }\\
{\footnotesize \ \ \ \ \ \ \ \ \ \ } \\
{\footnotesize Email : secil.demiray@bilecik.edu.tr, ftascan@ogu.edu.tr}}
\maketitle

\begin{abstract}
This paper is focused on quasi-periodic wave solutions of (3+1) generalized
BKP equation. Because of some difficulties in calculations of $N=3$ periodic
solutions, hardly ever has there been a study on these solutions by using
Rieamann theta function. In this study, we obtain one and two periodic wave
solutions as well as three periodic wave solutions for (3+1) generalized BKP
equation. Moreover we analyse the asymptotic behavior of the periodic wave
solutions tend to the known soliton solutions under a small amplitude limit.

\bigskip


\end{abstract}

\section{Introduction}

In recent years, the problem of finding exact solutions of partial
differential equations (PDE) is very popular for both mathematicians and
physcists. Because if we know the exact solutions of PDE's, they can help us
to understand complicated physical models. So, there are some successful
methods to obtain exact solutions such as Hirota's direct method \cite{[1]},
Lie symmetry method \cite{[2]}, B\"{a}cklund transformation method \cite{[3]}
and algebro geometric method \cite{[4]}.

In the late 1970's Novikov, Dubrovin, Mckean, Lax, Its, and Matveev et al.
developed the algebro geometric method to obtain quasi-periodic or
algebro-geometric solutions for many soliton equations \cite{[5],[8]}.
However this method involves complicated calculation. On the other hand,
Hirota's direct method is rather useful and direct approach to construct
multisoliton solutions.

In the 1980, Nakamura obtained the periodic wave solutions of the KdV and
the Boussinesq equations by means of Hirota's bilinear method \cite{[9],[10]}%
. Indeed this method has some advantages over algebro-geometric methods. We
can get explicit periodic wave solutions directly.

Recently, Fan and his collaborators have extended this method to investigate
the discrete Toda lattice \cite{[11]} , Cheng Z.,Hao X. studied on periodic
solution of (2+1) AKNS equation \cite{[12]}, Tian and Zhang obtained
periodic wave solutions by Riemann theta functions of some nonlinear\
differential equations and supersymmetric equations \cite{[13],[14]}, Lu and
Zhang studied on quasi periodic solutions of Jimbo-Miwa equation \cite{[15]}

Soliton equations possess nice mathematical features, e.g., elastic
interactions of solutions. Such equations contain the KdV equation, the
Boussinesq equation, the KP equation and the BKP equation, and they all have
multi-soliton solutions. Let us consider (3+1) dimensional generalized BKP
equation \cite{[16]}.%
\begin{equation}
u_{ty}-u_{xxxy}-3(u_{x}u_{y})_{x}+3u_{xz}=0  \tag{1.1}
\end{equation}

Now, in this paper we briefly introduce a Hirota bilinear form and the
Riemann theta function.Then after we apply the Hirota's bilinear method to
construct one, two and three periodic wave solutions to (3+1) generalized
BKP equation, respectively. We further use a limiting procedure to analyse
the asymptotic behavior of the periodic wave solutions in the last section.
It is rigorously shown that the periodic solutions tend to the well-known
soliton solutions under a certain limit.\vspace{12.7cm}

\section{The Bilinear Form and The Riemann Theta Functions}

In this section we introduce briefly bilinear form and some main points on
the Riemann theta functions. The Hirota bilinear method is powerful when
constructing exact solutions for nonlinear equations. Through the dependent
variable transformation $u=2(\ln f)_{x}$ , eq. (1.1) is written bilinear form%
\begin{equation}
(D_{y}D_{t}-D_{x}^{3}D_{y}+3D_{x}D_{z})f.f=0\ .  \tag{2.1}
\end{equation}%
Here $D$ is differential bilinear operator defined by%
\begin{equation}
\begin{array}{l}
D_{x}^{m}D_{y}^{n}D_{t}^{k}f(x,y,t).g(x,y,t)= \\ 
\\ 
(\partial _{x}-\partial _{x^{\prime }})^{m}(\partial _{y}-\partial
_{y^{\prime }})^{n}(\partial _{t}-\partial _{t^{\prime
}})^{k}f(x,y,t)g(x^{\prime },y^{\prime },t^{\prime })\mid _{x^{\prime
}=x,y^{\prime }=y,t^{\prime }=t}%
\end{array}
\tag{2.2}
\end{equation}%
and the operator has property for exponential functions namely%
\begin{equation}
D_{x}^{m}D_{y}^{n}D_{t}^{k}e^{\xi _{1}}e^{\xi _{2}}=(\alpha _{1}-\alpha
_{2})^{m}(\rho _{1}-\rho _{2})^{n}(\omega _{1}-\omega _{2})^{k}e^{\xi
_{1}+\xi _{2}}  \tag{2.3}
\end{equation}%
where $\xi _{i}=\alpha _{i}x+\rho _{i}y+\omega _{i}t+\delta _{i}$, $i=1,2.$
More general we can write following formula%
\begin{equation}
G(D_{x},D_{y},D_{t})e^{\xi _{1}}e^{\xi _{2}}=G(\alpha _{1}-\alpha _{2},\rho
_{1}-\rho _{2},\omega _{1}-\omega _{2})e^{\xi _{1}+\xi _{2}}  \tag{2.4}
\end{equation}%
where $G(D_{x},D_{y},D_{t})$ is \bigskip a polinomial about $D_{x},D_{y}$
and $D_{t}.\ $According to the Hirota bilinear theory, eq. (1.1) admits
one-soliton solution%
\begin{equation}
u_{1}=2\partial _{x}(\ln (1+e^{\eta }))  \tag{2.5}
\end{equation}%
where phase variable $\eta =\mu x+\nu y+\kappa z+\varpi t+\gamma $ ,
dispersion relation $\varpi =-3\frac{\mu \kappa }{\rho }$\bigskip +$\mu ^{3}$%
, $\mu ,\nu ,\kappa $ and $\gamma $ are constants.

Two-soliton solution%
\begin{equation}
u_{2}=2\partial _{x}(\ln (1+e^{\eta _{1}}+e^{\eta _{2}}+e^{\eta _{1}+\eta
_{2}+A_{12}}))  \tag{2.6}
\end{equation}%
with%
\begin{equation}
\begin{array}{c}
e^{A_{12}}=-\frac{(\nu _{1}-\nu _{2})(\varpi _{1}-\varpi _{2})-(\mu _{1}-\mu
_{2})^{3}(\nu _{1}-\nu _{2})+3(\mu _{1}-\mu _{2})(\kappa _{1}-\kappa _{2})}{%
(\nu _{1}+\nu _{2})(\varpi _{1}+\varpi _{2})-(\mu _{1}+\mu _{2})^{3}(\nu
_{1}+\nu _{2})+3(\mu _{1}+\mu _{2})(\kappa _{1}+\kappa _{2})}%
\end{array}
\tag{2.7}
\end{equation}%
\begin{equation}
\begin{array}{l}
\eta _{j}=\mu _{j}x+\nu _{j}y+\kappa _{j}z+\varpi _{j}t+\gamma _{j}\text{ ,\
\ \ \ \ \ \ }j=1,2 \\ 
\\ 
\varpi _{1}=-3\frac{\mu _{1}\kappa _{1}}{\rho _{1}}+\mu _{1}^{3},\ \varpi
_{2}=-3\frac{\mu _{2}\kappa _{2}}{\rho _{2}}+\mu _{2}^{3}%
\end{array}
\tag{2.8}
\end{equation}%
where $\mu _{j},\nu _{j},\kappa _{j}$ and $\gamma _{j}$ are arbitrary
constants.

Three-soliton solution%
\begin{equation}
u_{3}=2\partial _{x}(\ln (f))  \tag{2.9}
\end{equation}%
$f$ is written as%
\begin{equation}
\begin{array}{l}
f=1+e^{\eta _{1}}+e^{\eta _{2}}+e^{\eta _{3}}+e^{\eta _{1}+\eta _{2}+A_{12}}
\\ 
\ \ \ \ \ \ +e^{\eta _{1}+\eta _{3}+A_{13}}+e^{\eta _{2}+\eta
_{3}+A_{23}}+e^{\eta _{1}+\eta _{2}+\eta _{3}+A_{12}+A_{13}+A_{23}}%
\end{array}
\tag{2.10}
\end{equation}%
with%
\begin{equation}
\begin{array}{c}
e^{A_{ij}}=-\frac{(\nu _{i}-\nu _{j})(\varpi _{i}-\varpi _{j})-(\mu _{i}-\mu
_{j})^{3}(\nu _{i}-\nu _{j})+3(\mu _{i}-\mu _{j})(\kappa _{i}-\kappa _{j})}{%
(\nu _{i}+\nu _{j})(\varpi _{i}+\varpi _{j})-(\mu _{i}+\mu _{j})^{3}(\nu
_{i}+\nu _{j})+3(\mu _{i}+\mu _{j})(\kappa _{i}+\kappa _{j})}%
\end{array}
\tag{2.11}
\end{equation}%
\begin{equation}
\begin{array}{l}
\eta _{j}=\mu _{j}x+\nu _{j}y+\kappa _{j}z+\varpi _{j}t+\gamma _{j}\text{ ,\
\ \ \ \ \ }i,\text{\ }j=1,2,3\ \ ,i<j \\ 
\\ 
\varpi _{1}=-3\frac{\mu _{1}\kappa _{1}}{\rho _{1}}+\mu _{1}^{3},\ \varpi
_{2}=-3\frac{\mu _{2}\kappa _{2}}{\rho _{2}}+\mu _{2}^{3} \\ 
\\ 
\ \varpi _{3}=-3\frac{\mu _{3}\kappa _{3}}{\rho _{3}}+\mu _{3}^{3}%
\end{array}
\tag{2.12}
\end{equation}

In order to apply the Hirota's bilinear method to constact multi-periodic
wave solutions we consider a slightly generalized form of bilinear equation
(2.1). We look for our solution in the form 
\begin{equation}
\begin{array}{c}
u=u_{0}y+2(\ln \vartheta (\xi ))_{x} \\ 
\end{array}
\tag{2.13}
\end{equation}%
where $u_{0}y$ is a solution of (1.1) and phase variable $\xi =(\xi
_{1},...,\xi _{N})^{T}$ , $\xi _{i}=\alpha _{i}x+\rho _{i}y+k_{i}z+\omega
_{i}t+\delta _{i}$, $i=1,2..N.$

Substituting (2.13) into (1.1) and integration once respect to $x$ , we
obtain 
\begin{equation}
\begin{array}{c}
H(D_{x},D_{y},D_{z},D_{t},)=(D_{y}D_{t}+3D_{x}D_{z}-D_{x}^{3}D_{y}-3u_{0}D_{x}^{2}+c)\vartheta (\xi ).\vartheta (\xi )=0%
\end{array}
\tag{2.14}
\end{equation}%
where $c=c(y,z,t)$ is integration constant. For finding multiperiodic wave
solutions of (2.14), we consider the following multidimensional Riemann
theta function%
\begin{equation}
\vartheta (\xi ,\tau )=\sum_{n\in 
\mathbb{Z}
^{N}}e^{\pi i<\tau n,n>+2\pi i<\xi ,n>}  \tag{2.15}
\end{equation}%
where the integer value vector $n=(n_{1}...n_{N})^{T}\in 
\mathbb{Z}
^{N}$ and complex phase variables $\xi =(\xi _{1}...\xi _{N})^{T}\in 
\mathbb{C}
^{N}$ , for $N$ dimensional two vectors their inner product is defined by $%
<u,v>=u_{1}v_{1}+...+u_{N}v_{N}$ . Period matrix of theta function is -$%
i\tau =-i(\tau _{ij})$ which is positive definite and real-valued symmetric $%
N\times N$ matrix and can be considered as free parametres of theta
function. So the Fourier series (2.15) converges to a real valued function
and for make the theta function real valued in this paper we take $\tau $
imaginay matrix.

\begin{proposition}
The theta function $\vartheta (\xi ,\tau )$ has the periodic properties%
\begin{equation*}
\vartheta (\xi +1+\tau )=e^{-\pi i\tau -2\pi i\xi }\vartheta (\xi ,\tau )
\end{equation*}%
we regard the vectors $1$ and $\tau $ as a periods of the theta function $%
\vartheta (\xi ,\tau )$ with multipliers $1$ and $e^{-\pi i\tau -2\pi i\xi
}. $ Here $\tau $ is not a period of theta function $\vartheta (\xi ,\tau ),$
but it is the period of the functions $\partial _{\xi }^{2}\ln \vartheta
(\xi ,\tau )$, $\partial _{\xi }\ln [\vartheta (\xi +e,\tau )/\vartheta (\xi
+h,\tau )]$ and $\vartheta (\xi +e,\tau )\vartheta (\xi -e,\tau )/\vartheta
^{2}(\xi +h,\tau ).$
\end{proposition}

\section{One-periodic waves and asymptotic properties}

\subsection{Construct one periodic waves}

If we take $N=1,$ we obtain one-periodic solutions and our Riemann theta
function reduces following Fourier series%
\begin{equation}
\vartheta (\xi ,\tau )=\sum_{-\infty }^{\infty }e^{\pi in^{2}\tau +2\pi
in\xi }  \tag{3.1.1}
\end{equation}%
where the phase variable $\xi =\alpha x+\rho y+kz+\omega t+\delta $ and $%
\func{Im}(\tau )>0.$

\textbf{Theorem 1 }\textit{Assuming that }$\vartheta (\xi ,\tau )$\textit{\
is a Riemann theta function as \bigskip }$N=1$\textit{\ with }$\xi =\alpha
x_{1}+\rho x_{2}+...+\omega t+\delta $\textit{\ and }$\alpha ,\rho
,...,\omega ,\delta $\textit{\ \ satisfy the following system}%
\begin{equation}
\begin{array}{l}
\sum\limits_{n=-\infty }^{\infty }H(4n\pi i\alpha ,4n\pi i\rho ,...,4n\pi
i\omega )e^{2n^{2}\pi i\tau }=0\text{ \ }%
\end{array}
\tag{3.1.2}
\end{equation}%
\textit{\ }%
\begin{equation}
\begin{array}{l}
\sum\limits_{n=-\infty }^{\infty }H(2\pi i(2n-1)\alpha ,...,2\pi
i(2n-1)\omega )\ \ \ \ \ \ \ 
\end{array}
\tag{3.1.3}
\end{equation}%
\begin{equation*}
\begin{array}{c}
\times e^{(2n^{2}-2n+1)\pi i\tau }=0\ \ \ \ \ \ \ \ \ \ \ \ \ \ \ \ \ \ \ \
\ \ \ \ \ \ \ \ \ \ \ \ \ 
\end{array}%
\end{equation*}%
and \textit{the following expression}%
\begin{equation}
u=u_{0}y+2(\ln \vartheta (\xi ))_{x}  \tag{3.1.4}
\end{equation}%
\textit{is the one periodic wave solution of eq. (1.1). For the proof \ }%
\cite{[14]}.

According to the Theorem 1 $\alpha ,\rho ,k$ and $\omega $ should provide
the following system with (2.15)%
\begin{equation}
\begin{array}{l}
\widetilde{H}(0)=\dsum\limits_{n=-\infty }^{\infty }(-16\pi ^{2}n^{2}\rho
\omega -48\pi ^{2}n^{2}\alpha k-256\pi ^{4}n^{4}\rho \alpha ^{3} \\ 
\ \ \ \ \ \ \ \ \ \  \\ 
\ \ \ \ \ \ \ \ \ \ +48u_{0}\pi ^{2}n^{2}\alpha ^{2}+c)e^{2\pi in^{2}\tau }=0
\\ 
\\ 
\widetilde{H}(1)=\dsum\limits_{n=-\infty }^{\infty }(-4\pi
^{2}(2n-1)^{2}\rho \omega -12\pi ^{2}(2n-1)^{2}\alpha k-16\pi
^{4}(2n-1)^{4}\rho \alpha ^{3}\  \\ 
\\ 
\ \ \ \ \ \ \ \ \ \ +12\pi ^{2}u_{0}(2n-1)^{2}\alpha
^{2}+c)e^{(2n^{2}-2n+1)\pi i\tau }=0\ \ .%
\end{array}
\tag{3.1.5}
\end{equation}%
Our aim is solving this system about frequency$\ \omega $ and integration
constant $c$, namely%
\begin{equation}
\begin{pmatrix}
a_{11} & a_{12} \\ 
a_{21} & a_{22}%
\end{pmatrix}%
\begin{pmatrix}
\omega \\ 
c%
\end{pmatrix}%
=%
\begin{pmatrix}
b_{1} \\ 
b_{2}%
\end{pmatrix}%
.  \tag{3.1.6}
\end{equation}%
By introducing the notations as%
\begin{equation}
\begin{array}{l}
\lambda =e^{\pi i\tau }\text{ \ \ }a_{11}=\sum\limits_{n=-\infty }^{\infty
}-16\pi ^{2}n^{2}\rho \lambda ^{2n^{2}}\text{ \ \ } \\ 
\\ 
a_{12}=\sum\limits_{n=-\infty }^{\infty }\lambda ^{2n^{2}} \\ 
\\ 
a_{21}=\sum\limits_{n=-\infty }^{\infty }-4\pi ^{2}(2n-1)^{2}\rho \lambda
^{2n^{2}-2n+1} \\ 
\\ 
a_{22}=\sum\limits_{n=-\infty }^{\infty }\lambda ^{2n^{2}-2n+1} \\ 
\\ 
b_{1}=\sum\limits_{n=-\infty }^{\infty }(48\pi ^{2}n^{2}\alpha k+256\pi
^{4}n^{4}\rho \alpha ^{3}-48\pi ^{2}n^{2}\alpha ^{2}u_{0})\lambda ^{2n^{2}}%
\text{ \ \ \ \ } \\ 
\\ 
b_{2}=\sum\limits_{n=-\infty }^{\infty }(12\pi ^{2}(2n-1)^{2}\alpha k+16\pi
^{4}(2n-1)^{4}\rho \alpha ^{3} \\ 
\\ 
\ \ \ \ \ \ \ \ \ -12\pi ^{2}(2n-1)^{2}\alpha ^{2}u_{0})\lambda
^{2n^{2}-2n+1} \\ 
\end{array}
\tag{3.1.7}
\end{equation}%
we can easily solve this system and then we obtain a one-periodic wave
solution of Eq. (1.1)%
\begin{equation}
u=u_{0}y+2(\ln \vartheta (\xi ))_{x}  \tag{3.1.8}
\end{equation}%
where the parameters $\omega $ \ and $c$ \ are given by (3.1.7) but the
other parameters $\alpha ,\rho ,k,\delta ,\tau ,u_{0}$ are free.

\subsection{Asymptotic property of one periodic waves}

\textbf{Theorem 2 \ }\textit{If the vector }$(\omega ,c)^{T}$\textit{\ is a
solution of the system (3.1.6) and for the one-periodic wave solution
(3.1.8) we let }%
\begin{equation}
u_{0}=0,\ \text{\ \ \ }\alpha =\frac{\mu }{2\pi i},\text{ \ \ \ }\rho =\frac{%
\nu }{2\pi i},\text{ \ \ }k=\frac{\kappa }{2\pi i},\text{\ \ \ }\delta =%
\frac{\gamma -\pi i\tau }{2\pi i}  \tag{3.2.1}
\end{equation}%
\textit{where }$\mu ,\nu $\textit{\ and }$\gamma $\textit{\ are given (2.5).
Then we have following asymtotic properties}%
\begin{equation}
c\rightarrow 0,\text{ \ \ \ }\xi \rightarrow \frac{\eta -\pi i\tau }{2\pi i},%
\text{ \ \ \ \ \ \ \ }\vartheta (\xi ,\tau )\rightarrow 1+e^{\eta }\text{ \ }%
when\text{ \ }\lambda \rightarrow 0  \tag{3.2.2}
\end{equation}%
\textit{It implies that the one-periodic solution tends to the one-soliton
solution eq. (2.5) under a small amplitude limit}

\textbf{Proof }

The one-periodic wave solution (3.1.8) has two fundamental periods $1$ and $%
\tau $ in the phase variable $\xi $ . It's actually a kind of
one-dimensional cnoidal waves and speed parameter is given by%
\begin{equation}
\omega =\frac{b_{1}a_{22}-b_{2}a_{12}}{a_{11}a_{22}-a_{12}a_{21}}\ . 
\tag{3.2.3}
\end{equation}%
It has only one wave pattern for all time, and it can be viewed as a
parallel superposition of overlapping one-solitary waves, placed one period
apart

For consider asymptotic properties we have to find solution of system
(3.1.6) . Using eq. (3.1.7) coefficent matrix and the right-side vector of
system (3.1.6) are power series about $\lambda $ so its solution $(\omega
,c)^{T}$ also should be a series about $\lambda $%
\begin{equation*}
\begin{array}{l}
a_{11}=-32\pi ^{2}\rho \lambda ^{2}-128\pi ^{2}\rho \lambda ^{8}+... \\ 
\\ 
a_{12}=1+2\lambda ^{2}+2\lambda ^{8}+... \\ 
\\ 
a_{21}=-8\pi ^{2}\rho \lambda -72\pi ^{2}\rho \lambda ^{5}+... \\ 
\\ 
a_{22}=2\lambda +2\lambda ^{5}+... \\ 
\\ 
b_{1}=(96\pi ^{2}\alpha k+512\pi ^{4}\alpha ^{3}\rho -96u_{0}\pi ^{2}\alpha
^{2})\lambda ^{2} \\ 
\ \ +(384\pi ^{2}\alpha k+8192\pi ^{4}\alpha ^{3}\rho -384u_{0}\pi
^{2}\alpha ^{2})\lambda ^{8}+... \\ 
\\ 
b_{2}=(24\pi ^{2}\alpha k+32\pi ^{4}\alpha ^{3}\rho -24u_{0}\pi ^{2}\alpha
^{2})\lambda \\ 
\ \ +(216\pi ^{2}\alpha k+2592\pi ^{4}\alpha ^{3}\rho -216u_{0}\pi
^{2}\alpha ^{2})\lambda ^{5}+...%
\end{array}%
\end{equation*}

We can solve the system (3.1.6) via small parameter expansion method and\ we
obtain%
\begin{equation}
\begin{array}{l}
\omega =(-3\frac{\alpha k}{\rho }-4\pi ^{2}\alpha ^{3}+3u_{0}\frac{\alpha
^{2}}{\rho })+(96\pi ^{2}\alpha ^{3})\lambda ^{2}+(288\pi ^{2}\alpha
^{3})\lambda ^{4}+o(\lambda ^{4}) \\ 
\\ 
c=(384\pi ^{4}\rho \alpha ^{3})\lambda ^{2}+(2304\pi ^{4}\rho \alpha
^{3})\lambda ^{4}+o(\lambda ^{4})\ .%
\end{array}
\tag{3.2.4}
\end{equation}%
From Theorem 2 and (3.2.4), we have 
\begin{equation}
c\rightarrow 0,\text{ \ \ }\omega =-3\frac{\alpha k}{\rho }-4\pi ^{2}\alpha
^{3}\text{ }when\text{ \ }\lambda \rightarrow 0  \tag{3.2.5}
\end{equation}%
and substituting the relation (3.2.1) into (3.2.5) we obtain%
\begin{equation}
\varpi =2\pi i\omega =-3\frac{\mu \kappa }{\nu }\bigskip +\mu ^{3}\text{ .} 
\tag{3.2.6}
\end{equation}

The one-soliton solution of the (3+1) generalized BKP equation can be
obtained as a limit of the periodic solution (3.1.8). We can expand the
periodic function $\vartheta (\xi )$ in the following form 
\begin{eqnarray}
\vartheta (\xi ,\tau ) &=&\sum\limits_{-\infty }^{\infty }e^{\pi in^{2}\tau
+2\pi in\xi }  \TCItag{3.2.7} \\
&=&1+e^{\pi i\tau +2\pi i\xi }+e^{\pi i\tau -2\pi i\xi }+e^{4\pi i\tau +4\pi
i\xi }+...  \notag
\end{eqnarray}%
By using the transformation%
\begin{equation}
\begin{array}{l}
\xi \rightarrow \frac{\overset{\backsim }{\xi }-\pi i\tau }{2\pi i},\text{ \ 
}\lambda =e^{\pi i\tau } \\ 
\\ 
\vartheta (\xi ,\tau )=1+e^{\overset{\backsim }{\xi }}+\lambda ^{2}(e^{-%
\overset{\backsim }{\xi }}+e^{2\xi })+...%
\end{array}
\tag{3.2.8}
\end{equation}%
and when $\lambda \rightarrow 0$ we can write 
\begin{equation}
\vartheta (\xi ,\tau )=1+e^{\overset{\backsim }{\xi }}\ .  \tag{3.2.9}
\end{equation}%
According to one soliton solution $\overset{\backsim }{\xi }=\eta $ ,
therefore proof is completed.

\section{\protect\bigskip Two-periodic waves and asymptotic properties}

\subsection{Construct two-periodic waves}

We consider two-periodic wave solutions of Eq. (1.1) which are two
dimensional generalization of one-periodic wave solutions. Let's consider $%
N=2,$ and Riemann theta function takes the form 
\begin{equation}
\vartheta (\xi ,\tau )=\vartheta (\xi _{1,}\xi _{2,}\tau )=\sum\limits_{n\in 
\mathbb{Z}
^{2}}e^{\pi i<\tau n,n>+2\pi i<\xi ,n>}  \tag{4.1.1}
\end{equation}%
where $n=(n_{1,}n_{2})^{T}\in 
\mathbb{Z}
^{2},$ $\xi =(\xi _{1},\xi _{2})\in 
\mathbb{C}
^{2}$ , $\xi _{i}=\alpha _{i}x+\rho _{i}y+k_{i}z+\omega _{i}t+\delta _{i}$ \
, $i=1,2$ \ and $\ -i\tau $ is a positive definite and real-valued symmetric 
$2\times 2$ matrix which can take the form of 
\begin{equation}
\tau =%
\begin{pmatrix}
\tau _{11} & \tau _{12} \\ 
\tau _{12} & \tau _{22}%
\end{pmatrix}%
,\text{ \ \ \ \ }\func{Im}(\tau _{11})>0,\text{ }\func{Im}(\tau _{22})>0,%
\text{ }\tau _{11}\tau _{22}-\tau _{12}^{2}<0  \tag{4.1.2}
\end{equation}

\textbf{Theorem 3 \ }\textit{Assuming that }$\vartheta (\xi _{1,}\xi
_{2,}\tau )$\textit{\ is one Riemann theta function as }$N=2$\textit{\ with }%
$\xi _{i}=\alpha _{i}x+\rho _{i}y+k_{i}z+\omega _{i}t+\delta _{i}$\textit{\
and }$\alpha _{i},\rho _{i},k_{i},\omega _{i},\delta _{i}$\textit{\ \ ,}$%
i=1,2$\textit{\ satisfy the following system }%
\begin{equation}
\sum\limits_{n\in 
\mathbb{Z}
^{2}}H(2\pi i<2n-\theta _{j},\alpha >,...2\pi i<2n-\theta _{j},\omega >) 
\tag{4.1.3}
\end{equation}%
\begin{eqnarray*}
\times e^{\pi i[<\tau (n-\theta _{j}),n-\theta _{j}>+<\tau n,n>]} &=&0\text{
\ \ \ \ \ \ \ \ \ \ } \\
&&\text{\ \ \ \ \ \ \ \ }
\end{eqnarray*}%
\textit{where }$\theta _{j}=(\theta _{j}^{1},\theta _{j}^{2})^{T}$\textit{\
, }$\theta _{1}=(0,0)^{T},$\textit{\ }$\theta _{2}=(1,0)^{T},$\textit{\ }$%
\theta _{3}=(0,1)^{T},$\textit{\ }$\theta _{4}=(1,1)^{T}$\textit{\ , }$%
j=1,2,3,4$\textit{\ and the following expression}%
\begin{equation*}
u=u_{0}y+2(\ln \vartheta (\xi _{1,}\xi _{2},\tau ))_{x}
\end{equation*}%
\textit{is the two-periodic wave solution of Eq. (1.1). For the proof }\cite%
{[14]}

According to the Theorem 3 $\alpha _{i},\rho _{i},k_{i}$ and $\omega _{i}$
should provide the following system with (2.14)%
\begin{equation}
\begin{array}{l}
\sum\limits_{n\in 
\mathbb{Z}
^{2}}[-4\pi ^{2}<2n-\theta _{j},\rho ><2n-\theta _{j},\omega >-12\pi
^{2}<2n-\theta _{j},\alpha ><2n-\theta _{j},k> \\ 
\\ 
\ \ \ -16\pi ^{4}<2n-\theta _{j},\alpha >^{3}<2n-\theta _{j},\rho >+12\pi
^{2}u_{0}<2n-\theta _{j},\alpha >^{2} \\ 
\\ 
\ \ \ \ \ \ +c]\times e^{\pi i[<\tau (n-\theta _{j}),n-\theta _{j}>+<\tau
n,n>]}=0%
\end{array}
\tag{4.1.4}
\end{equation}%
where $j=1,2,3,4$. Our aim is solving this system namely%
\begin{equation}
X%
\begin{pmatrix}
\omega _{1} \\ 
\omega _{2} \\ 
u_{0} \\ 
c%
\end{pmatrix}%
=%
\begin{pmatrix}
b_{1} \\ 
b_{2} \\ 
b_{3} \\ 
b_{4}%
\end{pmatrix}
\tag{4.1.5}
\end{equation}%
where $X=(a_{ij})_{4\times 4}$ matrix.

By introducing the notation as 
\begin{equation}
\varepsilon _{j}=\lambda _{1}^{n_{1}^{2}+(n_{1}-\theta _{j}^{1})^{2}}\lambda
_{2}^{n_{2}^{2}+(n_{2}-\theta _{j}^{2})^{2}}\lambda
_{3}^{n_{1}n_{2}+(n_{1}-\theta _{j}^{1})(n_{2}-\theta _{j}^{2})}  \tag{4.1.6}
\end{equation}%
where 
\begin{equation}
\lambda _{1}=e^{\pi i\tau _{11}},\text{ \ \ \ }\lambda _{2}=e^{\pi i\tau
_{22}},\text{ \ \ \ }\lambda _{3}=e^{2\pi i\tau _{12}}\text{ \ \ and \ }%
j=1,2,3,4  \tag{4.1.7}
\end{equation}%
and

\begin{equation}
\begin{array}{l}
a_{j4}=\sum\limits_{n_{1,}n_{2}\in 
\mathbb{Z}
^{2}}\varepsilon _{j} \\ 
\\ 
a_{j3}=12\pi ^{2}\sum\limits_{n\in 
\mathbb{Z}
^{2}}<2n-\theta _{j},\alpha >^{2}\varepsilon _{j} \\ 
\\ 
a_{j2}=-4\pi ^{2}\sum\limits_{n\in 
\mathbb{Z}
^{2}}<2n-\theta _{j},\rho >(2n_{2}-\theta _{j}^{2})\varepsilon _{j} \\ 
\\ 
a_{j1}=-4\pi ^{2}\sum\limits_{n\in 
\mathbb{Z}
^{2}}<2n-\theta _{j},\rho >(2n_{1}-\theta _{j}^{1})\varepsilon _{j} \\ 
\\ 
b_{j}=\sum\limits_{n\in 
\mathbb{Z}
^{2}}12\pi ^{2}<2n-\theta _{j},\alpha ><2n-\theta _{j},k> \\ 
\\ 
+16\pi ^{4}<2n-\theta _{j},\alpha >^{3}<2n-\theta _{j},\rho >\varepsilon _{j}%
\end{array}
\tag{4.1.8}
\end{equation}%
we can solve this system and we obtain two-periodic wave solution as 
\begin{equation}
u=u_{0}y+2(\ln \vartheta (\xi _{1},\xi _{2},\tau ))_{x}  \tag{4.1.9}
\end{equation}%
where $\vartheta (\xi _{1},\xi _{2},\tau )$ and parameters $\omega
_{1},\omega _{2},u_{0},c$ are given by (4.1.1) and (4.1.5). The other $%
\alpha _{1},\alpha _{2},\rho _{1},\rho _{2},k_{1},k_{2},\tau _{11},\tau
_{12} $ and $\tau _{22\text{ }}$are arbitrary parameters .

We notice that the total number of unknown parameters $u_{0}\ $integration
constant $c$ $,\ $nonlinear frequency $\alpha _{i},\rho _{i},k_{i},\omega
_{i}$ and the term $\tau _{jk}=\tau _{kj}$ , $1\leq j,k\leq N$ \ is 
\begin{equation*}
\frac{1}{2}N(N+1)+4N+2\ .
\end{equation*}

\subsection{Asymptotic property of two periodic waves}

\textbf{Teorem 4 \ }\textit{If }$(\omega _{1},\omega _{2},u_{0},c)^{T}$%
\textit{\ is a solution of the system (4.1.5) and for the two-periodic wave
solution we take}%
\begin{equation}
\alpha _{j}=\frac{\mu _{j}}{2\pi i},\text{ \ }\rho _{j}=\frac{\nu _{j}}{2\pi
i},\ k_{j}=\frac{\kappa _{j}}{2\pi i}\text{ \ ,}\delta _{j}=\frac{\gamma
_{j}-\pi i\tau _{jj}}{2\pi i},\text{ \ }\tau _{12}=\frac{A_{12}}{2\pi i},%
\text{ \ }j=1,2  \tag{4.2.1}
\end{equation}%
\textit{where }$\mu _{j},\nu _{j,}\kappa _{j},\delta _{j}$\textit{\ and }$%
A_{12}$\textit{\ are given in Eq. (2.7) and (2.8) . Then we have the
following asymtotic relations}%
\begin{equation}
\begin{array}{l}
u_{0}\rightarrow 0,\text{ \ }c\rightarrow 0,\text{ \ }\xi _{j}\rightarrow 
\frac{\eta _{j}-\pi i\tau _{jj}}{2\pi i},\text{ \ }j=1,2 \\ 
\\ 
\vartheta (\xi _{1},\xi _{2},\tau )\rightarrow 1+e^{\eta _{1}}+e^{\eta
_{2}}+e^{\eta _{1}+\eta _{2}+A_{12}}\text{ \ \ as \ \ }\lambda _{1},\lambda
_{2}\rightarrow 0%
\end{array}
\tag{4.2.2}
\end{equation}%
\textit{\ That means the two-periodic solution tends to the two-solion
solution under a small amplitude limit. }

\textbf{Proof \ }The Riemann theta function is\textbf{\ }%
\begin{equation}
\vartheta (\xi _{1},\xi _{2},\tau )=\sum\limits_{n\in 
\mathbb{Z}
^{2}}e^{\pi i<\tau n,n>+2\pi i<\xi ,n>}  \tag{4.2.3}
\end{equation}%
Let's expand this function%
\begin{eqnarray}
&&\sum\limits_{n_{1},n_{2}\in 
\mathbb{Z}
^{2}}e^{2\pi i(\xi _{1}n_{1}+\xi _{2}n_{2})+\pi i[n_{1}(\tau _{11}n_{1}+\tau
_{12}n_{2})+n_{2}(\tau _{12}n_{1}+\tau _{22}n_{2})]}  \TCItag{4.2.4} \\
&=&1+e^{2\pi i\xi _{1}+\pi i\tau _{11}}+e^{-2\pi i\xi _{1}+\pi i\tau
_{11}}+...  \notag
\end{eqnarray}%
and if we take $\xi _{j}\rightarrow \frac{\overset{\sim }{\xi }_{j}-\pi
i\tau _{jj}}{2\pi i}$ in Eq. (4.2.4) we have%
\begin{equation}
\vartheta (\xi _{1},\xi _{2},\tau )=1+e^{\overset{\sim }{\xi _{1}}}+e^{%
\overset{\sim }{\xi _{2}}}+e^{\overset{\sim }{\xi _{1}}+\overset{\sim }{\xi
_{2}+2\pi i\tau _{12}}}+\lambda _{1}^{2}e^{-\overset{\sim }{\xi _{1}}%
}+\lambda _{2}^{2}e^{-\overset{\sim }{\xi _{2}}}+...  \tag{4.2.5}
\end{equation}%
where $\lambda _{1}=e^{\pi i\tau _{11}},$ \ \ \ $\lambda _{2}=e^{\pi i\tau
_{22}}$ and $\lambda _{1},\lambda _{2}\rightarrow 0$ 
\begin{equation}
\vartheta (\xi _{1},\xi _{2},\tau )=1+e^{\overset{\sim }{\xi _{1}}}+e^{%
\overset{\sim }{\xi _{2}}}+e^{\overset{\sim }{\xi _{1}}+\overset{\sim }{\xi
_{2}+2\pi i\tau _{12}}}\ .  \tag{4.2.6}
\end{equation}%
According to the two soliton solution (2.6) we can write 
\begin{equation}
\tau _{12}=\frac{A_{12}}{2\pi i}  \tag{4.2.7}
\end{equation}%
For solving system (4.1.5) we can expand each funcion into a series with $%
\lambda _{1}$ and $\lambda _{2}$%
\begin{equation}
\begin{array}{c}
X=X_{0}+X_{1}\lambda _{1}+X_{2}\lambda _{2}+X_{11}\lambda
_{1}^{2}+X_{22}\lambda _{2}^{2} \\ 
+X_{12}\lambda _{1}\lambda _{2}+o(\lambda _{1}^{k},\lambda _{2}^{j})\ \ ,\ \
\ k+l\geq 2.%
\end{array}
\tag{4.2.8}
\end{equation}

and%
\begin{equation}
\begin{array}[t]{l}
\begin{pmatrix}
\omega _{1} \\ 
\omega _{2} \\ 
u_{0} \\ 
c%
\end{pmatrix}%
=%
\begin{pmatrix}
\omega _{1}^{0} \\ 
\omega _{2}^{0} \\ 
u_{0}^{0} \\ 
c^{0}%
\end{pmatrix}%
+%
\begin{pmatrix}
\omega _{1}^{1} \\ 
\omega _{2}^{1} \\ 
u_{0}^{1} \\ 
c^{1}%
\end{pmatrix}%
\lambda _{1}+%
\begin{pmatrix}
\omega _{1}^{2} \\ 
\omega _{2}^{2} \\ 
u_{0}^{2} \\ 
c^{2}%
\end{pmatrix}%
\lambda _{2}+%
\begin{pmatrix}
\omega _{1}^{3} \\ 
\omega _{2}^{3} \\ 
u_{0}^{3} \\ 
c^{3}%
\end{pmatrix}%
\lambda _{1}^{2} \\ 
\\ 
+%
\begin{pmatrix}
\omega _{1}^{4} \\ 
\omega _{2}^{4} \\ 
u_{0}^{4} \\ 
c^{4}%
\end{pmatrix}%
\lambda _{2}^{2}+%
\begin{pmatrix}
\omega _{1}^{5} \\ 
\omega _{2}^{5} \\ 
u_{0}^{5} \\ 
c^{5}%
\end{pmatrix}%
\lambda _{1}\lambda _{2}+o(\lambda _{1}^{k}\lambda _{2}^{l})\text{ , \ \ }%
k+l\geq 2%
\end{array}
\tag{4.2.9}
\end{equation}

Substituting these equations into the (4.1.5), we obtain%
\begin{equation}
\begin{array}{l}
c=(384\pi ^{4}\alpha _{1}^{3}\rho _{1})\lambda _{1}^{2}+(384\pi ^{4}\alpha
_{2}^{3}\rho _{2})\lambda _{2}^{2}+o(\lambda _{1},\lambda _{2})\ \ \  \\ 
\\ 
\omega _{1}=(-3\frac{\alpha _{1}k_{1}}{\rho _{1}}-4\pi ^{2}\alpha _{1}^{3}+3%
\frac{\alpha _{1}^{2}}{\rho _{1}}u_{0}^{0})+(3\frac{\alpha _{1}^{2}}{\rho
_{1}}u_{0}^{1})\lambda _{1}+(3\frac{\alpha _{1}^{2}}{\rho _{1}}u_{0}^{2}%
\text{ })\lambda _{2} \\ 
\\ 
\ \ \ \ \ \ +o(\lambda _{1},\lambda _{2}) \\ 
\\ 
\omega _{2}=(-3\frac{\alpha _{2}k_{2}}{\rho _{2}}-4\pi ^{2}\alpha _{2}^{3}+3%
\frac{\alpha _{2}^{2}}{\rho _{2}}u_{0}^{0})+(3\frac{\alpha _{2}^{2}}{\rho
_{2}}u_{0}^{1})\lambda _{1}+(3\frac{\alpha _{2}^{2}}{\rho _{2}}u_{0}^{2}%
\text{ })\lambda _{2} \\ 
\ \ \ \ \ \ \ +o(\lambda _{1},\lambda _{2}).%
\end{array}
\tag{4.2.10}
\end{equation}%
If we choose $u_{0}^{0}=0$ , and $(\lambda _{1},\lambda _{2})\rightarrow
(0,0)$, we can find%
\begin{eqnarray}
u_{0} &=&o(\lambda _{1},\lambda _{2})\rightarrow 0\text{ \ , \ \ }%
c\rightarrow 0  \TCItag{4.2.11} \\
\omega _{1} &=&-3\frac{\alpha _{1}k_{1}}{\rho _{1}}-4\pi ^{2}\alpha _{1}^{3}
\notag \\
\omega _{2} &=&-3\frac{\alpha _{2}k_{2}}{\rho _{2}}-4\pi ^{2}\alpha _{2}^{3}.
\notag
\end{eqnarray}%
According to the Theorem 4, we obtain%
\begin{equation}
\begin{array}{l}
\varpi _{1}=-\frac{3\mu _{1}\kappa _{1}}{\nu _{1}}+\mu _{1}^{3}\text{ \ , \ }%
\varpi _{2}=-\frac{3\mu _{2}\kappa _{2}}{\nu _{2}}+\mu _{2}^{3}\text{ , }%
c\rightarrow 0\text{\ } \\ 
\\ 
\text{when }\ u_{0}=o(\lambda _{1},\lambda _{2})\rightarrow 0\text{ \ }.%
\end{array}
\tag{4.2.12}
\end{equation}%
and when solving the system we obtain 
\begin{equation}
\begin{array}{c}
\lambda _{3}=-\frac{(\nu _{1-}\nu _{2})(\varpi _{1}-\varpi _{2})-(\mu
_{1}-\mu _{2})^{3}(\nu _{1}-\nu _{2})+3(\mu _{1}-\mu _{2})(\kappa
_{1}-\kappa _{2})}{(\nu _{1+}\nu _{2})(\varpi _{1}+\varpi _{2})-(\mu
_{1}+\mu _{2})^{3}(\nu _{1}+\nu _{2})+3(\mu _{1}+\mu _{2})(\kappa
_{1}+\kappa _{2})}%
\end{array}
\tag{4.2.13}
\end{equation}%
That means just by solving system we can obtain $e^{A_{12}},$this is
alternative proof for $\tau _{12}=\frac{A_{12}}{2\pi i}$

From (4.2.12), we conclude that the two-periodic solution tends to the two
soliton solution as $\lambda _{1},\lambda _{2}\rightarrow 0.$

\section{Three-periodic waves and asymptotic properties}

We consider three-periodic wave solutions of Eq. \ref{1.1}. Let's consider $%
N=3$, and Riemann theta function takes the form 
\begin{equation}
\vartheta (\xi ,\tau )=\vartheta (\xi _{1,}\xi _{2,}\xi _{3,}\tau
)=\sum\limits_{n\in 
\mathbb{Z}
^{3}}e^{\pi i<\tau n,n>+2\pi i<\xi ,n>}  \tag{5.1.1}
\end{equation}%
where $n=(n_{1,}n_{2},n_{3})^{T}\in 
\mathbb{Z}
^{3},$ $\xi =(\xi _{1},\xi _{2},\xi _{3})\in 
\mathbb{C}
^{3}$ , $\xi _{i}=\alpha _{i}x+\rho _{i}y+k_{i}z+\omega _{i}t+\delta _{i}$ \
, $i=1,2,3$ \ and $\ -i\tau $ is a positive definite and real-valued
symmetric $3\times 3$ matrix which can take the form of 
\begin{equation}
\tau =%
\begin{pmatrix}
\tau _{11} & \tau _{12} & \tau _{13} \\ 
\tau _{12} & \tau _{22} & \tau _{23} \\ 
\tau _{13} & \tau _{23} & \tau _{33}%
\end{pmatrix}%
,\text{ \ \ \ \ }\func{Im}(\tau _{jk})>0,\ \text{ }j=k=1,2,3  \tag{5.1.2}
\end{equation}

\textbf{Theorem 5 \ }\textit{Assuming that }$\vartheta (\xi _{1,}\xi
_{2,}\xi _{3},\tau )$\textit{\ is one Riemann theta function as }$N=3$%
\textit{\ with }$\xi _{i}=\alpha _{i}x+\rho _{i}y+k_{i}z+\omega _{i}t+\delta
_{i}$\textit{\ and }$\alpha _{i},\rho _{i},k_{i},\omega _{i},\delta _{i}$%
\textit{\ \ ,}$i=1,2,3$\textit{\ satisfy the following system }%
\begin{equation}
\begin{array}{c}
\sum\limits_{n\in 
\mathbb{Z}
^{3}}H(2\pi i<2n-\theta _{j},\alpha >,...2\pi i<2n-\theta _{j},\omega >) \\ 
e^{\pi i[<\tau (n-\theta _{j}),n-\theta _{j}>+<\tau n,n>]}=0 \\ 
\end{array}
\tag{5.1.3}
\end{equation}%
\textit{where }$\theta _{j}=(\theta _{j}^{1},\theta _{j}^{2},\theta
_{j}^{3})^{T},\ \theta _{1}=(0,0,0)^{T},\ \theta _{2}=(0,0,1)^{T},\ \theta
_{3}=(0,1,0)^{T},\ \theta _{4}=(0,1,1)^{T},\ \theta _{5}=(1,0,0)^{T},\
\theta _{6}=(1,0,1)^{T},\ \theta _{7}=(1,1,0)^{T},\ \theta
_{8}=(1,1,1)^{T},\ j=1,..,8\ $\textit{and the following expression}

\begin{equation}
u=u_{0}y+2(\ln \vartheta (\xi _{1},\xi _{2},\xi _{3},\tau ))_{x}  \tag{5.1.4}
\end{equation}%
\textit{is the three-periodic wave solution.}

\bigskip

\textbf{Proof. \ \ }Substituting (5.1.1) into bilinear equation $%
H(D_{x},D_{y},D_{z},D_{t})\ $and using yhe property (2.4), we have following
result%
\begin{equation}
\begin{array}{l}
H(D_{x},D_{y},D_{z},D_{t})\vartheta (\xi _{1},\xi _{2},\xi _{3},\tau
).\vartheta (\xi _{1},\xi _{2},\xi _{3},\tau ) \\ 
\\ 
=\sum\limits_{m,n\in 
\mathbb{Z}
^{3}}H(2\pi i<n-m,\alpha >,...2\pi i<n-m,\omega >) \\ 
\\ 
\ \ \ \ \ e^{2\pi i<\xi ,m+n>+\pi i(<\tau m,m>+<\tau n,n>)} \\ 
\\ 
=\sum\limits_{m^{\prime }\in 
\mathbb{Z}
^{3}}\{\sum\limits_{n\in 
\mathbb{Z}
^{3}}H(2\pi i<2n-m^{\prime },\alpha >,...2\pi i<2n-m^{\prime },\omega >) \\ 
\\ 
\ \ \ \ \ \ e^{\pi i(<\tau (n-m^{\prime }),n-m^{\prime }>+<\tau
n,n>)}\}e^{2\pi i<\xi ,m^{\prime }>} \\ 
\\ 
=\sum\limits_{m^{\prime }\in 
\mathbb{Z}
^{3}}\hat{H}(m_{1}^{\prime },m_{2}^{\prime },m_{3}^{\prime })e^{2\pi i<\xi
,m^{\prime }>} \\ 
\\ 
=\sum\limits_{m^{\prime }\in 
\mathbb{Z}
^{3}}\hat{H}(m^{\prime })e^{2\pi i<\xi ,m^{\prime }>},\ m^{\prime }=m+n%
\end{array}
\tag{5.1.5}
\end{equation}%
Shifting index n as $n^{\prime }=n-\delta _{ij},\ j=1,2,3$ we can compute
that%
\begin{equation}
\begin{array}{l}
\hat{H}(m^{\prime })=\hat{H}(m_{1}^{\prime },m_{2}^{\prime },m_{3}^{\prime })
\\ 
\\ 
=\sum\limits_{n\in 
\mathbb{Z}
^{3}}H(2\pi i<2n-m^{\prime },\alpha >,...2\pi i<2n-m^{\prime },\omega >) \\ 
\\ 
\ \ \ \ e^{\pi i(<\tau (n-m^{\prime }),n-m^{\prime }>+<\tau n,n>)} \\ 
\\ 
=\sum\limits_{n\in 
\mathbb{Z}
^{3}}H(2\pi i\sum\limits_{i=1}^{3}[2n_{i}^{\prime }-(m_{i}^{\prime }-2\delta
_{ij})]\alpha _{i},...,2\pi i\sum\limits_{i=1}^{3}[2n_{i}^{\prime
}-(m_{i}^{\prime }-2\delta _{ij})]\omega _{i}) \\ 
\\ 
\ \ \ \ e^{\pi i\sum\limits_{i,k=1}^{3}[(n_{i}^{\prime }+\delta
_{ij})(n_{k}^{\prime }+\delta _{kj})+(m_{i}^{\prime }-n_{i}^{\prime }-\delta
_{ij})(m_{k}^{\prime }-n_{k}^{\prime }-\delta _{kj})]\tau _{ik}} \\ 
=\left\{ 
\begin{array}{c}
\hat{H}(m_{1}^{\prime }-2,m_{2}^{\prime },m_{3}^{\prime })e^{2\pi
i(m_{1}^{\prime }-1)\tau _{11}+2\pi i(m_{2}^{\prime }\tau
_{12}+m_{3}^{\prime }\tau _{13})\ \ \ },\ j=1 \\ 
\hat{H}(m_{1}^{\prime },m_{2}^{\prime }-2,m_{3}^{\prime })e^{2\pi
i(m_{2}^{\prime }-1)\tau _{22}+2\pi i(m_{1}^{\prime }\tau
_{12}+m_{3}^{\prime }\tau _{13})\ \ \ },\ j=2 \\ 
\hat{H}(m_{1}^{\prime },m_{2}^{\prime },m_{3}^{\prime }-2)e^{2\pi
i(m_{3}^{\prime }-1)\tau _{33}+2\pi i(m_{1}^{\prime }\tau
_{11}+m_{2}^{\prime }\tau _{12})\ \ \ },\ j=3%
\end{array}%
\right.%
\end{array}
\tag{5.1.6}
\end{equation}%
which implies that if%
\begin{equation}
\hat{H}(m_{1}^{\prime },m_{2}^{\prime },m_{3}^{\prime })=0  \tag{5.1.7}
\end{equation}%
hold for all combinations of $m_{1}^{\prime }=0,1,\ m_{2}^{\prime }=0,1,\
m_{3}^{\prime }=0,1$ , then all $\hat{H}(m_{1}^{\prime },m_{2}^{\prime
},m_{3}^{\prime })=0,\ m_{i}^{\prime }\in 
\mathbb{Z}
^{3}\ (i=1,2,3)\ $and $\delta _{ij\text{ }}$ representing Kronecker's delta.
If we require%
\begin{equation}
\begin{array}{l}
\hat{H}(m^{\prime })=\sum\limits_{n\in 
\mathbb{Z}
^{3}}H(2\pi i<2n-\theta _{j},\alpha >,...2\pi i<2n-\theta _{j},\omega >) \\ 
\\ 
\ \ \ \ \ \ \ \ \ \ \ \ e^{\pi i(<\tau (n-\theta _{j}),n-\theta _{j}>+<\tau
n,n>)}%
\end{array}
\tag{5.1.8}
\end{equation}%
where $\theta _{j}=(\theta _{j}^{1},\theta _{j}^{2},\theta _{j}^{3})^{T}\ $%
and $\theta _{1}=(0,0,0)^{T},\ \theta _{2}=(0,0,1)^{T},\ \theta
_{3}=(0,1,0)^{T},\ \theta _{4}=(0,1,1)^{T},\ \theta _{5}=(1,0,0)^{T},\
\theta _{6}=(1,0,1)^{T},\ \theta _{7}=(1,1,0)^{T},\ \theta
_{8}=(1,1,1)^{T},\ j=1,..,8,$ we can obtain three-periodic wave solutions.

According to the Theorem 5 $\alpha _{i},\rho _{i},k_{i}$ and $\omega _{i}$
should provide the following system with (2.14)

\begin{equation}
\begin{array}{l}
\sum\limits_{(n_{1,}n_{2},n_{3})\in 
\mathbb{Z}
^{3}}[-4\pi ^{2}<2n-\theta _{j},\rho ><2n-\theta _{j},\omega > \\ 
\\ 
\ \ \ \ \ \ \ -12\pi ^{2}<2n-\theta _{j},\alpha ><2n-\theta _{j},k>-16\pi
^{4}<2n-\theta _{j},\alpha >^{3}<2n-\theta _{j},\rho > \\ 
\\ 
\ \ \ \ \ \ \ +12\pi ^{2}u_{0}<2n-\theta _{j},\alpha >^{2}+c]\times e^{\pi
i[<\tau (n-\theta _{j}),n-\theta _{j}>+<\tau n,n>]}=0%
\end{array}
\tag{5.1.9}
\end{equation}%
where $j=1,...,8$. Our aim is solving this system namely%
\begin{equation}
X(\omega _{1},\omega _{2},\omega _{3},k_{1},k_{2},k_{3},u_{0},c)^{T}=b 
\tag{5.1.10}
\end{equation}%
where $X=(a_{ij})_{8\times 8}$ matrix and $%
b=(b_{1},b_{2},b_{3},b_{4},b_{5},b_{6},b_{7},b_{8})$.

By introducing the notation as%
\begin{equation}
\begin{array}{l}
\varepsilon _{j}=\sum\limits_{(n_{1,}n_{2},n_{3})\in 
\mathbb{Z}
^{3}}e^{\pi i[<\tau (n-\theta _{j}),n-\theta _{j}>+<\tau n,n>]} \\ 
\\ 
\ \ \ \ =\lambda _{1}^{n_{1}^{2}+(n_{1}-\theta _{j}^{1})^{2}}\lambda
_{2}^{n_{2}^{2}+(n_{2}-\theta _{j}^{2})^{2}}\lambda
_{3}^{n_{3}^{2}+(n_{3}-\theta _{j}^{3})^{2}} \\ 
\\ 
\ \ \ \ \ \ \ \lambda _{12}^{n_{1}n_{2}+(n_{1}-\theta _{j}^{1})(n_{2}-\theta
_{j}^{2})}\lambda _{13}^{n_{1}n_{3}+(n_{1}-\theta _{j}^{1})(n_{3}-\theta
_{j}^{3})}\lambda _{23}^{n_{2}n_{3}+(n_{2}-\theta _{j}^{2})(n_{3}-\theta
_{j}^{3})}%
\end{array}
\tag{5.1.11}
\end{equation}%
where

\begin{equation}
\begin{array}{l}
\lambda _{1}=e^{\pi i\tau _{11}},\text{\ }\lambda _{2}=e^{\pi i\tau _{22}},\
\lambda _{3}=e^{\pi i\tau _{33}} \\ 
\\ 
\lambda _{12}=e^{2\pi i\tau _{12}},\ \lambda _{13}=e^{2\pi i\tau _{13}},\
\lambda _{23}=e^{2\pi i\tau _{23}} \\ 
\\ 
j=1,..,8%
\end{array}
\tag{5.1.12}
\end{equation}%
and

\begin{equation}
\begin{array}{l}
a_{j8}=\sum\limits_{n\in 
\mathbb{Z}
^{3}}\varepsilon _{j} \\ 
\\ 
a_{j7}=\sum\limits_{n\in 
\mathbb{Z}
^{3}}12\pi ^{2}<2n-\theta _{j},\alpha >^{2}\varepsilon _{j} \\ 
\\ 
a_{j6}=\sum\limits_{n\in 
\mathbb{Z}
^{3}}-12\pi ^{2}<2n-\theta _{j},\alpha >(2n_{3}-\theta _{j}^{3})\varepsilon
_{j} \\ 
\\ 
a_{j5}=\sum\limits_{n\in 
\mathbb{Z}
^{3}}-12\pi ^{2}<2n-\theta _{j},\alpha >(2n_{2}-\theta _{j}^{2})\varepsilon
_{j} \\ 
\\ 
a_{j4}=\sum\limits_{n\in 
\mathbb{Z}
^{3}}-12\pi ^{2}<2n-\theta _{j},\alpha >(2n_{1}-\theta _{j}^{1})\varepsilon
_{j} \\ 
\\ 
a_{j3}=\sum\limits_{n\in 
\mathbb{Z}
^{3}}-4\pi ^{2}<2n-\theta _{j},\rho >(2n_{3}-\theta _{j}^{3})\varepsilon _{j}
\\ 
\\ 
a_{j2}=\sum\limits_{n\in 
\mathbb{Z}
^{3}}-4\pi ^{2}<2n-\theta _{j},\rho >(2n_{2}-\theta _{j}^{2})\varepsilon _{j}
\\ 
\\ 
a_{j1}=\sum\limits_{n\in 
\mathbb{Z}
^{3}}-4\pi ^{2}<2n-\theta _{j},\rho >(2n_{1}-\theta _{j}^{1})\varepsilon _{j}
\\ 
\\ 
b_{j}=\sum\limits_{n\in 
\mathbb{Z}
^{3}}16\pi ^{4}<2n-\theta _{j},\alpha >^{3}<2n-\theta _{j},\rho >\varepsilon
_{j}%
\end{array}
\tag{5.1.13}
\end{equation}%
we can solve this system and we obtain three-periodic wave solution as 
\begin{equation*}
u=u_{0}y+2(\ln \vartheta (\xi _{1},\xi _{2},\xi _{3},\tau ))_{x}
\end{equation*}%
where $\vartheta (\xi _{1},\xi _{2},\xi _{3},\tau )$ and parameters $\omega
_{1},\omega _{2},\omega _{3},k_{1},k_{2},k_{3},u_{0},c\ $are given by
(5.1.1) and (5.1.10). The other $\alpha _{1},\alpha _{2},\alpha _{3},\rho
_{1},\rho _{2},\rho _{3},\tau _{11},\tau _{22},\tau _{33},\tau _{12},\tau
_{13}$ and $\tau _{23\text{ }}$are arbitrary parameters .

\subsection{Asymptotic property of three periodic waves}

\textbf{Teorem 6 \ }\textit{If }$(\omega _{1},\omega _{2},\omega
_{3},k_{1},k_{2},k_{3},u_{0},c)^{T}$\textit{\ is a solution of the system
(5.1.10) and for the three-periodic wave solution we take}%
\begin{equation}
\begin{array}{l}
\alpha _{j}=\frac{\mu _{j}}{2\pi i},\text{ \ }\rho _{j}=\frac{\nu _{j}}{2\pi
i},\ k_{j}=\frac{\kappa _{j}}{2\pi i}\text{ \ ,}\delta _{j}=\frac{\gamma
_{j}-\pi i\tau _{jj}}{2\pi i}, \\ 
\\ 
\text{\ }\tau _{ij}=\frac{A_{ij}}{2\pi i},\text{ \ }i,j=1,2,3,\ i<j%
\end{array}
\tag{5.2.1}
\end{equation}%
\textit{where }$\mu _{j},\nu _{j,}\kappa _{j},\delta _{j}$\textit{\ and }$%
A_{ij}$\textit{\ are given in Eq. (2.11) and (2.12) . Then we have the
following asymtotic relations}%
\begin{equation}
\begin{array}{l}
u_{0}\rightarrow 0,\text{ \ }c\rightarrow 0,\text{ \ }\xi _{j}\rightarrow 
\frac{\eta _{j}-\pi i\tau _{jj}}{2\pi i},\text{ \ }j=1,2,3 \\ 
\\ 
\vartheta (\xi _{1},\xi _{2},\xi _{3,}\tau )\rightarrow 1+e^{\eta
_{1}}+e^{\eta _{2}}+e^{\eta _{3}}+e^{\eta _{1}+\eta _{2}+A_{12}} \\ 
\\ 
+e^{\eta _{1}+\eta _{3}+A_{13}}+e^{\eta _{2}+\eta _{3}+A_{23}}+e^{\eta
_{1}+\eta _{2}+\eta _{3}+A_{12}+A_{13}+A_{23}}\text{\ } \\ 
\\ 
\text{\ as \ \ }\lambda _{1},\lambda _{2},\lambda _{3}\rightarrow 0.%
\end{array}
\tag{5.2.2}
\end{equation}%
\textit{\ That means the three-periodic solution tends to the three-solion
solution under a small amplitude limit. }

\textbf{Proof \ }The Riemann theta function is\textbf{\ }%
\begin{equation}
\vartheta (\xi _{1},\xi _{2},\xi _{3},\tau )=\sum\limits_{n\in 
\mathbb{Z}
^{3}}e^{\pi i<\tau n,n>+2\pi i<\xi ,n>}  \tag{5.2.3}
\end{equation}%
Let's expand this function%
\begin{equation}
\begin{array}{l}
=\sum\limits_{n_{1},n_{2},n_{3}\in 
\mathbb{Z}
^{3}}e^{2\pi i(\xi _{1}n_{1}+\xi _{2}n_{2}+\xi _{3}n_{3})+\pi i[\tau
_{11}n_{1}^{2}+\tau _{22}n_{2}^{2}+\tau _{33}n_{3}^{2}+2n_{1}n_{2}\tau
_{12}+2n_{1}n_{3}\tau _{13}+2n_{2}n_{3}\tau _{23})]} \\ 
\\ 
=1+e^{2\pi i\xi _{1}+\pi i\tau _{11}}+e^{-2\pi i\xi _{1}+\pi i\tau
_{11}}+e^{2\pi i\xi _{2}+\pi i\tau _{22}}+e^{-2\pi i\xi _{2}+\pi i\tau _{22}}
\\ 
\\ 
+e^{2\pi i\xi _{3}+\pi i\tau _{33}}+e^{-2\pi i\xi _{3}+\pi i\tau
_{33}}+e^{\pi i\tau _{11}+\pi i\tau _{22}+2\tau _{12}+2\pi i\xi _{1}+2\pi
i\xi _{2}}+...%
\end{array}
\tag{5.2.4}
\end{equation}%
and if we take $\xi _{j}\rightarrow \frac{\overset{\sim }{\xi }_{j}-\pi
i\tau _{jj}}{2\pi i}$ in Eq. (5.2.4) we have%
\begin{equation}
\begin{array}{l}
\vartheta (\xi _{1},\xi _{2},\tau )=1+e^{\overset{\sim }{\xi _{1}}}+e^{%
\overset{\sim }{\xi _{2}}}+e^{\overset{\sim }{\xi _{3}}}+e^{\overset{\sim }{%
\xi _{1}}+\overset{\sim }{\xi _{2}}+2\pi i\tau _{12}}+e^{\overset{\sim }{\xi
_{1}}+\overset{\sim }{\xi _{3}}+2\pi i\tau _{13}} \\ 
\\ 
+e^{\overset{\sim }{\xi _{2}}+\overset{\sim }{\xi _{3}}+2\pi i\tau _{23}}+e^{%
\overset{\sim }{\xi _{1}}+\overset{\sim }{\xi _{2}}+\overset{\sim }{\xi _{3}}%
+2\pi i\tau _{12}+2\pi i\tau _{13}+2\pi i\tau _{23}}+\lambda _{1}^{2}e^{-%
\overset{\sim }{\xi _{1}}}+\lambda _{2}^{2}e^{-\overset{\sim }{\xi _{2}}} \\ 
\\ 
+\lambda _{3}^{2}e^{-\overset{\sim }{\xi _{2}}}+\lambda _{1}^{2}\lambda
_{2}^{2}e^{-\overset{\sim }{\xi _{1}}-\overset{\sim }{\xi _{2}}+2\pi i\tau
_{12}}+...%
\end{array}
\tag{5.2.5}
\end{equation}%
where $\lambda _{1}=e^{\pi i\tau _{11}},$\ $\lambda _{2}=e^{\pi i\tau
_{22}},\ \lambda _{3}=e^{\pi i\tau _{33}}\ \ \ $and $\lambda _{1},\lambda
_{2},\lambda _{3}\rightarrow 0$%
\begin{equation}
\begin{array}{l}
\vartheta (\xi _{1},\xi _{2},\xi _{3},\tau )=1+e^{\overset{\sim }{\xi _{1}}%
}+e^{\overset{\sim }{\xi _{2}}}+e^{\overset{\sim }{\xi _{3}}}+e^{\overset{%
\sim }{\xi _{1}}+\overset{\sim }{\xi _{2}}+2\pi i\tau _{12}}+e^{\overset{%
\sim }{\xi _{1}}+\overset{\sim }{\xi _{3}}+2\pi i\tau _{13}} \\ 
\\ 
+e^{\overset{\sim }{\xi _{2}}+\overset{\sim }{\xi _{3}}+2\pi i\tau _{23}}+e^{%
\overset{\sim }{\xi _{1}}+\overset{\sim }{\xi _{2}}+\overset{\sim }{\xi _{3}}%
+2\pi i\tau _{12}+2\pi i\tau _{13}+2\pi i\tau _{23}}%
\end{array}
\tag{5.2.6}
\end{equation}%
According to the three-soliton solution (2.9) we can write 
\begin{equation}
\tau _{12}=\frac{A_{12}}{2\pi i},\ \tau _{13}=\frac{A_{13}}{2\pi i},\ \tau
_{23}=\frac{A_{23}}{2\pi i}  \tag{5.2.7}
\end{equation}%
For solving system (5.1.10) we can expand each funcion into a series with $%
\lambda _{1},\lambda _{2}\ $and $\lambda _{3}$%
\begin{equation}
\begin{array}{l}
X=X_{0}+X_{1}\lambda _{1}+X_{2}\lambda _{2}+X_{3}\lambda _{3}+X_{4}\lambda
_{1}^{2}+X_{5}\lambda _{2}^{2}+X_{6}\lambda _{3}^{2} \\ 
\\ 
\ \ \ +X_{7}\lambda _{1}\lambda _{2}+X_{8}\lambda _{1}\lambda
_{3}+X_{9}\lambda _{2}\lambda _{3}+...\ 
\end{array}
\tag{5.2.8}
\end{equation}%
and we obtain \ \ \ 

\begin{equation}
\begin{array}{l}
c=(384\pi ^{4}\alpha _{1}^{3}\rho _{1})\lambda _{1}^{2}+(384\pi ^{4}\alpha
_{2}^{3}\rho _{2})\lambda _{2}^{2}+(384\pi ^{4}\alpha _{3}^{3}\rho
_{3})\lambda _{3}^{2}+o(\lambda _{1}^{i},\lambda _{2}^{j},\lambda _{3}^{k})\
\ ,\ i+j+k\geq 3\ \ \  \\ 
\\ 
\omega _{1}=(-3\frac{\alpha _{1}k_{1}^{(0)}}{\rho _{1}}-4\pi ^{2}\alpha
_{1}^{3}+3\frac{\alpha _{1}^{2}}{\rho _{1}}u_{0}^{(0)})+(-3\frac{\alpha
_{1}k_{1}^{(1)}}{\rho _{1}}+3\frac{\alpha _{1}^{2}}{\rho _{1}}%
u_{0}^{(1)})\lambda _{1}+(-3\frac{\alpha _{1}k_{1}^{(2)}}{\rho _{1}}+3\frac{%
\alpha _{1}^{2}}{\rho _{1}}u_{0}^{(2)}\text{ })\lambda _{2} \\ 
\\ 
\ \ \ \ \ \ +(-3\frac{\alpha _{1}k_{1}^{(3)}}{\rho _{1}}+3\frac{\alpha
_{1}^{2}}{\rho _{1}}u_{0}^{(3)}\text{ })\lambda _{3}+... \\ 
\\ 
\omega _{2}=(-3\frac{\alpha _{2}k_{2}^{(0)}}{\rho _{2}}-4\pi ^{2}\alpha
_{2}^{3}+3\frac{\alpha _{2}^{2}}{\rho _{2}}u_{0}^{(0)})+(-3\frac{\alpha
_{2}k_{2}^{(1)}}{\rho _{2}}+3\frac{\alpha _{2}^{2}}{\rho _{2}}%
u_{0}^{(1)})\lambda _{1}+(-3\frac{\alpha _{2}k_{2}^{(2)}}{\rho _{2}}+3\frac{%
\alpha _{2}^{2}}{\rho _{2}}u_{0}^{(2)}\text{ })\lambda _{2} \\ 
\\ 
\ \ \ \ \ \ \ +(-3\frac{\alpha _{2}k_{2}^{(3)}}{\rho _{2}}+3\frac{\alpha
_{2}^{2}}{\rho _{2}}u_{0}^{(3)}\text{ })\lambda _{3}+... \\ 
\\ 
\omega _{3}=(-3\frac{\alpha _{3}k_{3}^{(0)}}{\rho _{3}}-4\pi ^{2}\alpha
_{3}^{3}+3\frac{\alpha _{3}^{2}}{\rho _{3}}u_{0}^{(0)})+(-3\frac{\alpha
_{3}k_{3}^{(1)}}{\rho _{3}}+3\frac{\alpha _{3}^{2}}{\rho _{3}}%
u_{0}^{(1)})\lambda _{1}+(-3\frac{\alpha _{3}k_{3}^{(2)}}{\rho _{3}}+3\frac{%
\alpha _{3}^{2}}{\rho _{3}}u_{0}^{(2)}\text{ })\lambda _{2} \\ 
\\ 
\ \ \ \ \ \ \ \ +(-3\frac{\alpha _{3}k_{3}^{(3)}}{\rho _{3}}+3\frac{\alpha
_{3}^{2}}{\rho _{3}}u_{0}^{(3)}\text{ })\lambda _{3}+...%
\end{array}
\tag{5.2.9}
\end{equation}%
where we expand the notations as follows%
\begin{equation}
\begin{array}{l}
k_{i}=k_{i}^{(0)}+k_{i}^{(1)}\lambda _{1}+k_{i}^{(2)}\lambda
_{2}+k_{i}^{(3)}\lambda _{3}+k_{i}^{(11)}\lambda
_{1}^{2}+k_{i}^{(22)}\lambda _{2}^{2} \\ 
\\ 
\ \ \ +k_{i}^{(33)}\lambda _{3}^{2}+k_{i}^{(12)}\lambda _{1}\lambda
_{2}+k_{i}^{(13)}\lambda _{1}\lambda _{3}+k_{i}^{(23)}\lambda _{2}\lambda
_{3}+..\ i=1,2,3%
\end{array}
\tag{5.2.10}
\end{equation}%
and parameters $\omega _{i},c\ $and $u_{0}\ $are similar to (5.2.10).

If we choose $u_{0}^{0}=0$ , and $(\lambda _{1},\lambda _{2},\lambda
_{3})\rightarrow (0,0,0)$, we can find%
\begin{equation}
\begin{array}{c}
u_{0}\rightarrow 0\text{ \ , \ \ }c\rightarrow 0 \\ 
\\ 
\omega _{1}=-3\frac{\alpha _{1}k_{1}}{\rho _{1}}-4\pi ^{2}\alpha _{1}^{3} \\ 
\\ 
\omega _{2}=-3\frac{\alpha _{2}k_{2}}{\rho _{2}}-4\pi ^{2}\alpha _{2}^{3} \\ 
\\ 
\omega _{3}=-3\frac{\alpha _{3}k_{3}}{\rho _{3}}-4\pi ^{2}\alpha _{3}^{3}%
\end{array}
\tag{5.2.11}
\end{equation}%
According to the Theorem 6, we obtain%
\begin{equation}
\begin{array}{l}
\varpi _{1}=-\frac{3\mu _{1}\kappa _{1}}{\nu _{1}}+\mu _{1}^{3}\text{ \ , \ }%
\varpi _{2}=-\frac{3\mu _{2}\kappa _{2}}{\nu _{2}}+\mu _{2}^{3}\text{ } \\ 
\\ 
\varpi _{3}=-\frac{3\mu _{3}\kappa _{3}}{\nu _{3}}+\mu _{3}^{3}\text{, \ \ \ 
}c\rightarrow 0\text{\ } \\ 
\\ 
\text{when }\ u_{0}=o(\lambda _{1},\lambda _{2},\lambda _{3})\rightarrow 0%
\text{ \ }.%
\end{array}
\tag{5.2.12}
\end{equation}

From (5.2.12), we conclude that the three-periodic solution tends to the
three soliton solution as $\lambda _{1},\lambda _{2},\lambda _{3}\rightarrow
0$\bigskip

\subsection{Conclusion}

In this paper, we have obtained the one, two and three periodic wave
solutions of the (3+1) generalized BKP equation, \ by using Hirota's
bilinear method and the Riemann theta functions. Moreover, we have shown
that they can be reduced to classical solitons, under a small amplitude
limit.

The results can be extended to the case $N\geq 4$ but when solving the
system we need more unknown parameters so there is certain difficulties in
the calculation and it is still open problem for us .

\subsection{Acknowledments}

This study was supported by the Eskisehir Osmangazi University (ESOGU BAP:
201419A206).

\end{document}